\documentclass[twocolumn]{aastex62}
\usepackage{textcomp}
\usepackage{graphicx}
\usepackage{here}
\usepackage{hhline}
\usepackage{longtable}
\usepackage{comment}
\usepackage{CJK}

\def\red<#1>{\textcolor{red}{#1}}

\received{February 11, 2019}
\revised{March 6, 2019}
\accepted{March 12, 2019}

\shortauthors{Goda \& Matsuo}

\begin{document}
\begin{CJK*}{UTF8}{min}

\title{Multiple Populations of Extrasolar Gas Giants}

\author{Shohei Goda}
\affil{\rm Department of Earth and Space Science, Graduate School of Science, Osaka University, 1-1, Machikaneyamacho, Toyonaka, Osaka 560-0043, Japan; goda@iral.ess.sci.osaka-u.ac.jp, matsuo@iral.ess.sci.osaka-u.ac.jp}

\author{Taro Matsuo}
\affil{\rm Department of Earth and Space Science, Graduate School of Science, Osaka University, 1-1, Machikaneyamacho, Toyonaka, Osaka 560-0043, Japan; goda@iral.ess.sci.osaka-u.ac.jp, matsuo@iral.ess.sci.osaka-u.ac.jp}
\affil{\rm NASA Ames Research Center, Space Science and Astrobiology Division, Moffett Field, CA 94035, USA}

\begin{abstract}

There are two planetary formation scenarios: core accretion and gravitational disk instability. Based on the fact that gaseous objects are preferentially observed around metal-rich host stars, most extra-solar gaseous objects discovered to date are thought to have been formed by core accretion. Here, we present 569 samples of gaseous planets and brown dwarfs found in 485 planetary systems that span three mass regimes with boundary values at 4 and 25 Jupiter-mass masses through performing cluster analyses of these samples regarding the host-star metallicity, after minimizing the impact of the selection effect of radial-velocity measurement on the cluster analysis. The larger mass is thought to be the upper mass limit of the objects that were formed during the planetary formation processes. In contrast, the lower mass limit appears to reflect the difference between planetary formation processes around early-type and G-type stars; disk instability plays a greater role in the planetary formation process around early-type stars than that around G-type stars. Population with masses between 4 and 25 Jupiter masses that orbit early-type stars comprise planets formed not only via the core-accretion process but also via gravitational disk instability because the population preferentially orbits metal-poor stars or is independent of the host-star metallicity. Therefore, it is essential to have a hybrid scenario for the planetary formation of the diverse systems.

\end{abstract}

\vspace{1cm}


\section{Introduction} \label{sec:introduction}

The discussion of planetary formation was developed decades ago for the solar system \citep{1985prpl.conf.1100H}. Two representative formation scenarios for Jupiter have been proposed: core accretion \citep{1974Icar...22..416P, 1980PThPh..64..544M, 1996Icar..124...62P} and disk instability \citep{1951PNAS...37....1K, 1997Sci...276.1836B, 2002Sci...298.1756M}. In theory, the two planet-formation processes have different dependences on the proto-planetary-disk metallicity - defined as the ratio of the number-density of metals to hydrogen atoms - and on the planetary mass \citep[e.g.,][]{2007ApJ...662.1282M}. In the core accretion model, a proto-planet core easily grows to the critical core mass before the disk gas dissipates. This occurs because the disk metallicity reflects the building materials available for the core \citep{2004ApJ...616..567I, 2012A&A...541A..97M}. In fact, since the first planet orbiting a normal star was discovered \citep{1995Natur.378..355M}, large-sized radial-velocity observations have revealed that, although the metallicities of stars hosting smaller planets - such as Neptune-like planets and super-Earths - are significantly lower than those of stars orbited by extrasolar gas giants \citep{2011arXiv1109.2497M, 2015AJ....149...14W}, the gas giants preferentially orbit metal-rich stars \citep[e.g.,][]{2003A&A...398..363S, 2005ApJ...622.1102F}. Because the central star and its surrounding proto-planetary disk are formed from the same molecular cloud, according to the primordial hypothesis, most gas giants are thought to have formed via core accretion. Regarding the planetary mass, gas giants with planetary masses up to 30 $\rm M_J$ are potentially formed via core accretion \citep[e.g.,][]{2007ApJ...667..557T, 2016ApJ...823...48T}, where $\rm M_J$ represents the Jupiter mass. The number of gas giants more massive than a few $\rm M_J$ gradually decreases as the planetary mass increases \citep[e.g.,][]{2009A&A...501.1161M}.

For the disk instability scenario, the relationship between disk metallicity and disk-instability-induced planetary formation has been studied theoretically. There are reports of negative correlation \citep{2006ApJ...636L.149C, 2007Arizona}, a very weak positive correlation \citep{2007ApJ...661L..77M}, and no correlation \citep{2002ApJ...567L.149B} with the metallicities of the stars hosting the observed planets. Although the mass distribution of gas giants formed via the disk instability still remains an open question, a lower limit may exist on the masses of the disk-instability-induced planets \citep{2007ApJ...662.1282M}. On the other hand, direct imaging of the extrasolar planets orbiting HR8799, Formalhaut, and beta Pictoris \citep{2008Sci...322.1348M, 2008Sci...322.1345K, 2010Sci...329...57L} has confirmed the existence of outer planets, which can be explained better by the disk-instability scenario rather than by extended core accretion with migration or planet-planet scattering \citep{2009ApJ...707...79D}. Additionally, \cite{2016ApJ...825...63C} showed that it is difficult to form gaseous objects beyond 10 au via pebble accretion. In fact, the dynamical masses of the four planets orbiting HR 8799, which were constrained using high-precision astrometric measurements \citep{2018AJ....156..192W}, were consistent with those predicted from their infrared fluxes; assuming that the four planets follows the hot-start evolutionary track \citep{2003A&A...402..701B}. Thus, two populations of planets that have originated from the two different planet-formation mechanisms may exist.

Several discrepancies existed between the planetary distributions predicted by the core-accretion model and the observation results. The paucity of planets due to the rapid gas accretion of the core-accretion model \citep{2004ApJ...604..388I, 2014A&A...567A.121D} is not consistent with the existence of abundant gas dwarfs close to the host stars, as revealed by the Kepler data \citep[e.g.,][]{2018ApJS..235...38T}. In addition, \cite{2018ApJ...869L..34S} proved that the Saturn-mass planets beyond the snow line around late-type stars, which were detected by the microlensing method, are much more plentiful than the mass distributions predicted by the core-accretion models \citep{2004ApJ...604..388I, 2009A&A...501.1161M}. Furthermore, the pebble accretion model that plays an important role in formation of proto-planetary core \citep{2010A&A...520A..43O, 2012A&A...544A..32L} has certain drawbacks. For example, \cite{2018MNRAS.480.4338L} showed that the pebble accretion rate is largely limited by the fast radial drift speed of mm--cm-sized pebbles because of the gas drag; this rate is insufficient for the formation of cores, which leads to runaway gas accretion. Note that pebble accretion might form planetary systems similar to the solar system, including gas giants, when the pebble accretion rate is enhanced by icy pebbles beyond the snow line \citep{2016ApJ...825...63C}. \cite{2018ApJ...865...30C} tuned the parameters of the pebble accretion model to the frequencies of planets with masses ranging from 0.01 to 100 $\rm M_J$ and to the semi-major axes of 0.01 to 10 au derived by several observational surveys. The tuned model successfully generated short-period gas dwarfs, but it failed to reproduce hot Jupiters. Regarding the inner migration, \cite{2014MNRAS.445..479C} showed that no gas giants survived in their N-body simulations because of rapid inner type-I migration induced by the saturated co-rotation torque of the proto-planet core. Previous studies on population synthesis have applied optimistic migration for the survival of gas giants. Thus, the current core-accretion models need to be improved to bridge the gap between the observation results and tht eheories. Another formation scenario such as disk instability may be also required.

A hybrid scenario for planetary formation was first discussed by \cite{2007A&A...464..779R} and \cite{2007ApJ...662.1282M}; they focused on the distribution of host-star metallicities and planet masses. Subsequently, \cite{2017A&A...603A..30S} and \cite{2018AJ....156..221N} investigated the relation between the host-star metallicities and planet masses for the samples hosted by various spectral types of stars,> and showed that the gas giants are divided into two regions separated by a boundary mass of 4 $\rm M_J$. They have interpreted the two populations as outcomes originating from the two planetary formation mechanisms; gas giants lighter than 4 $\rm M_J$ are core-accreted planets, while those more massive than 4 $\rm M_J$ are formed through the disk instability. However, gas giants are expected to have formed differently and evolved around the various spectral types of stars in the core-accretion mechanism, given that the proto-planetary disk mass varies with the host-star masses \citep{2005ApJ...626.1045I}. The disk lifetime, which largely impacts the core-accretion scenario, also depends on the host-star masses \citep{2015A&A...576A..52R}. In addition, outer gas giants, which are much larger than 4 $\rm M_J$ and detected by direct imaging, preferentially orbit early-type stars. Thus, it is preferable to perform statistical analysis for each spectral type of host star. Conversely, \cite{2018ApJ...853...37S} restricted the samples to the gaseous objects orbiting G-type stars, and showed that there is a transition between 4 and 10 $\rm M_J$ instead of a clear boundary at 4 $\rm M_J$. However, only seven samples had masses ranging from 4 to 25 $\rm M_J$, and it was difficult to explain why the planetary formation changed in that mass regime. In addition, although all the previous studies have concluded that the two planetary formation mechanisms are separated by a boundary mass (with or without a transition), the two mechanisms may coexist in the same mass regime.

Based on this background, we classified a large number of extrasolar gaseous samples discovered by radial velocity and investigated how the planetary distribution changes depending on the spectral type of host star. We report on 569 samples of gaseous planets and brown dwarfs, which we have divided into three mass regimes separated by boundary masses at 4 and 25 $\rm M_J$. We determined that the host-star metallicities of the samples with masses ranging from 0.3 to 4 $\rm M_J$ and from 4 to 25 $\rm M_J$ around G-type stars are continuously the same, and these host-star metallicities are higher than those for the samples more massive than 25 $\rm M_J$. Given that a massive core-inducing runaway gas accretion onto the core prior to the disk lifetime tends to be formed in the disk with high metallicity \citep{2004ApJ...616..567I, 2012A&A...541A..97M}, the upper mass limit for the core-accreted planets is approximately 25 $\rm M_J$. The fact that the samples more massive than 25 $\rm M_J$ around early-type stars is deficit has also been confirmed. Thus, the upper boundary mass is thought to represent the upper mass limits for the planetary objects that can be formed around G-type and early-type stars. However, the lower boundary mass represents a difference between the planetary formation processes around G-type and early-type stars; disk instability has a more important role in planetary formation around more early-type stars as compared with the planetary formation around G-type stars. In this paper, we define G-type and early-type stars as those with masses ranging from 0.8 to 1.3 $\rm M_{\odot}$ and more massive than 1.3 $\rm M_{\odot}$, respectively. Note that we have constructed the samples used for this study so that the impact of radial-velocity selection effects on the statistical analysis are minimized. 

This paper is organized as follows. In Section \ref{sec:method}, first, we explain how we constructed the samples used for our statistical analyses, introducing the ``common-biased samples," which we have selected to minimize the impact of the difference between selection effects in metal-rich and -poor regions on the analyses. In Section \ref{sec:results}, we derive the boundary metallicity that divides the samples into two regions for which the distributions of planetary masses differ the most. By applying a Gaussian-mixture model, we also show that the samples are divided into three mass regimes, which arise from the difference between the distributions of gaseous objects orbiting G-type and early-type stars. In Section \ref{sec:discussion}, by comparing the results of the statistical analysis with the two planetary formation models, we discuss the upper mass limit for gas giants formed via bottom-up planetary formation, and we consider whether disk-instability-induced planetary formation occurs.

\section{Method} \label{sec:method}

In this section, we explain how we constructed the samples, determining the boundary between gas dwarfs - such as Neptune-like planets - and gaseous giants, and we show how to deal with the selection effect of the radial-velocity measurements by which the samples used for this study were detected.

\subsection{Preparation of Samples} \label{subsec:preparation}

\begin{figure}[t]
\begin{center}
\includegraphics[width=8.5cm]{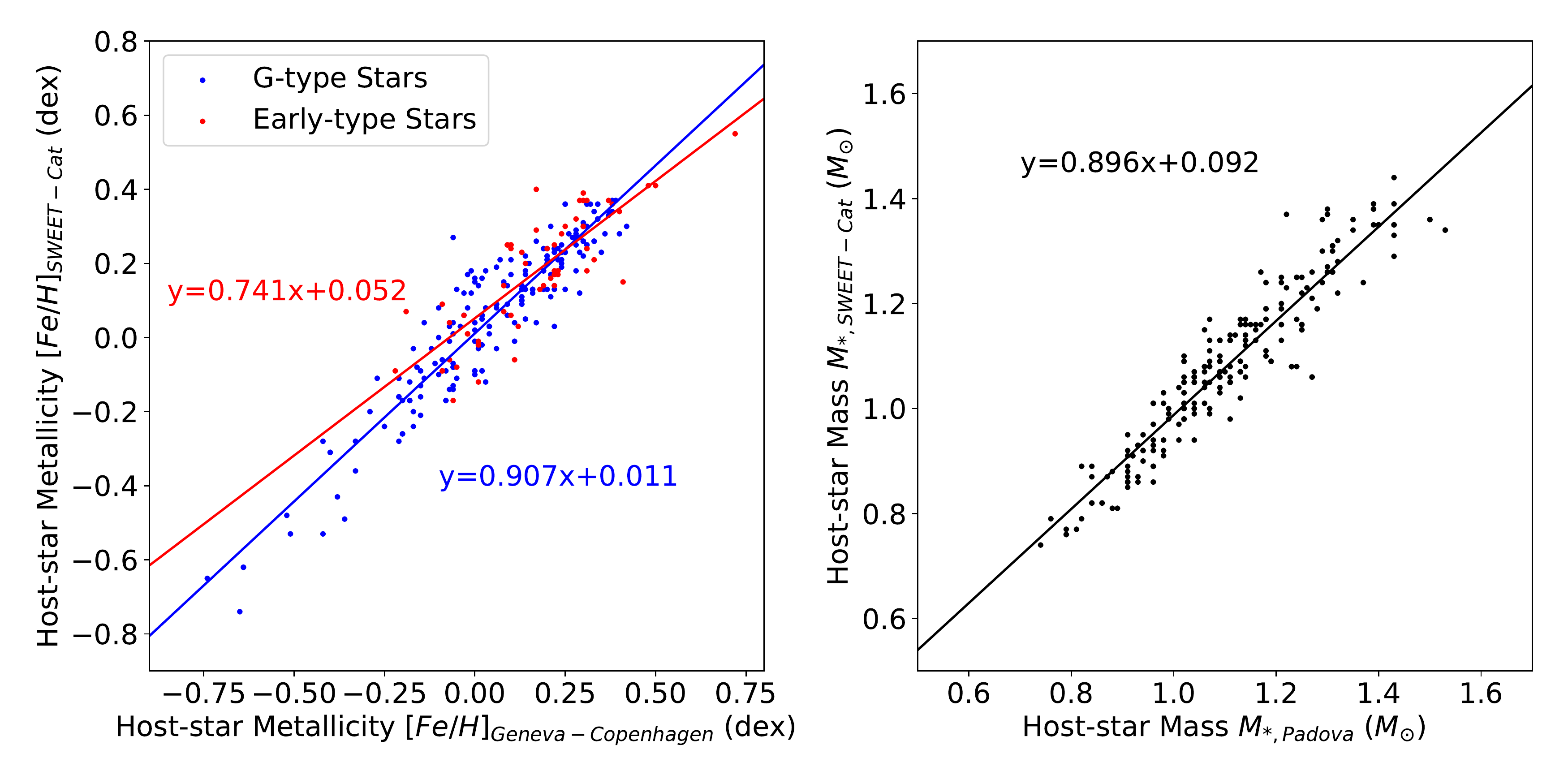}
\caption{$Left:$ Host-star metallicity correlations between the SWEET-Cat and Geneva-Copenhagen catalogs for G-type (blue) and early-type stars (red). There are 211 G-type stars and 60 early-type stars common to both catalogs. The variables y and x in the linear regression equations represent the host-star metallicities from the SWEET-Cat and Geneva-Copenhagen catalogs, respectively. Note that in this paper we define the G-type and early-type stars, respectively, defined as those with masses ranging from 0.8 to 1.3 $\rm M_{\odot}$ and those more massive than 1.3 $\rm M_{\odot}$. $Right:$ Host-star mass correlations between the SWEET-Cat catalog and Padova database. There are 239 samples common to both catalogs.}
\label{fig:linear}
\end{center}
\end{figure}

To determine the distributions of masses and orbital properties for samples orbiting various host-star of various metallicities, we collected extrasolar gaseous objects that have been discovered by radial-velocity observations for which the lower limit of companion mass, the semi-major axis and the eccentricity can be determined precisely. We term these objects as the ``original samples." Essentially, we selected the original samples from those labeled ``Radial Velocity" in the ``detection method" column of the Extrasolar Planet Encyclopedia catalog as of the end of June 2018 \citep{2011A&A...532A..79S}. The radial velocities of the host stars orbited by the original samples, as well as the orbital periods and eccentricities of the original samples, were collected from publications that reported planet detection and updates on masses and orbital parameters. The planet name, semi-amplitude of radial velocity, orbital period, eccentricity, lower limit of object mass, semi-major axis, and its reference for each original sample are compiled in Table \ref{tab:planet}. We referred to the SWEET-Cat catalog for the metallicity and mass of each host star \citep{2013A&A...556A.150S, 2018A&A...620A..58S}; this catalog presents uniformly derived stellar parameters for the planet-host stars \citep{2008A&A...487..373S}. For some of the original samples that are not listed in the SWEET-Cat catalog, we used the metallicities and masses compiled in the Geneva-Copenhagen catalog \citep{2011A&A...530A.138C} and Padova database \citep{2000A&AS..141..371G}, and calibrated them by using regression lines for G-type and early-type stars. We determined the regression lines from the metallicity and mass correlations among the SWEET-Cat, Geneva-Copenhagen catalogs, and Padova database to minimize measurement biases for host-star metallicities and masses (see Figure \ref{fig:linear}). Note that, samples that were derived not by the uniform method \citep{2008A&A...487..373S} exist in the SWEET-Cat catalog, the metallicities and host-star masses of the samples were directly used. This is because there is a good correlation between the metallicities/masses derived using the uniform method and those listed in publications. Finally, we calibrated 34 and three samples that, respectively, orbit G-type and early-type stars in terms of the host-star metallicity. Because the stellar masses were thus revised, according to Kepler's third law (Equation (\ref{equ:a})), we used the new stellar mass $M_*$ in $\rm M_{\odot}$, to re-calculate the semi-major axis of planet $a$ as below:
\begin{equation}
\label{equ:a}
a \approx 9.996\times10^{-1}P^\frac{2}{3}M_{*}^\frac{1}{3}\ \ [{\rm au}] ,
\end{equation}
where $P$ is the orbital period in years. The lower limit to the companion mass $M_p$ is also calculated as below \citep{2008ApJ...677.1324T}:
\begin{equation}
\label{equ:Mp}
M_{p}{\sin}i \approx 4.919{\times}10^{-3}P^\frac{1}{3}(1-e^2)^\frac{1}{2}KM_{*}^\frac{2}{3}\ \ [{\rm M_J}] ,
\end{equation}
where $P$ is in days, $e$ is the orbital eccentricity, and $K$ is the velocity semi-amplitude of the star in $\rm ms^{-1}$.

As indicators of the selection effect, we extracted the measurement accuracy and duration of observations for the radial-velocity measurement from the publications listed in Table \ref{tab:star}. The duration of observation and host-star mass provide the upper limit to the semi-major axis of a detectable companion for each radial-velocity measurement from Equation (\ref{equ:a_max}). Using the derived maximum semi-major axis, host-star mass, and measurement accuracy, we derived the lower mass limit for detectable companion from Equation (\ref{equ:Mp_min}).

\subsection{Boundary between Gas Giants and Neptune-like Planets} \label{subsec:boundary}

We extracted only gaseous objects from all the samples in the Extrasolar Planet Encyclopedia catalog to remove the impact of low-mass samples - such as Neptune-mass planets (gas dwarfs) and super-Earths - for this analysis. We determined the boundary mass between the gas-giant and gas-dwarf objects from the perspective of both theory and observation. According to a previous study \citep{2004ApJ...604..388I}, gas-dwarf objects, consisting primarily of heavy-core objects such as Neptune and Uranus, have the potential to grow to the extent allowed by the core-building materials inside their semi-major axes. This growth occurs via giant impacts in the inner region of the disk after the disk gas dissipates. However, this core growth is limited by scattering from the heavy core, which increases with greater distance from the central star. Therefore, the mass of a gas-dwarf object reaches a maximum at the semi-major axis, where the scattering effect begins to limit the core growth. Assuming the ratio of collision-to-ejection probabilities for the heavy core to be 0.1 and the core density to be 1 $\rm gcm^{-3}$, we obtain an upper mass limit for a gas-dwarf object $M_{gd}|_{max}$ as below:
\begin{equation}
M_{gd}|_{max} \approx 0.37\left(\frac{\eta_{ice}}{4}\right)^\frac{3}{4}\left(\frac{f_d}{10}\right)^\frac{3}{4}\ \  [\rm M_J] ,
\end{equation}
where $\eta_{ice}$ is the dust surface-density enhancement factor due to ice condensation and $f_d$ if the scaling factor of the dust surfave density relative to the Minimum Solar Nebulae (MMSN) model value. Considering a factor of the solid-angle average of $\sin{i}$, $\pi/4$, $M_{p}{\sin}i$ corresponding to the upper mass limit on a gas-dwarf object is estimated to be approximately 0.3 $\rm M_J$ for the dust surface density of 10 times the MMSN model value.

On the other hand, \cite{2017A&A...604A..83B} showed that there is a transition between the gas dwarfs and gas giants in the planetary mass-radius relation and the transition occurs at a mass of about 120 $\rm M_{\oplus}$, and then concluded that planets with masses larger than 120 $\rm M_{\oplus}$ mostly correspond to gas giants (i.e., hydrogen-helium-dominated planets) through comparing the theoretical calculations. Considering the factor of the solid-angle average, $M_{p}{\sin}i$ corresponding to the mass at which the transition occurs is approximately 0.3 $\rm M_J$. Based on these considerations, we used 0.3 $\rm M_J$ in this study as the boundary mass between gas giants and gas dwarfs. For this study, we considered 569 samples of gaseous planets or substellar objects belonging to 485 planetary systems.

\subsection{Common-biased Samples} \label{subsec:common}

\begin{figure}
\begin{minipage}[t]{1\hsize}
\begin{center}
\includegraphics[width=8.5cm]{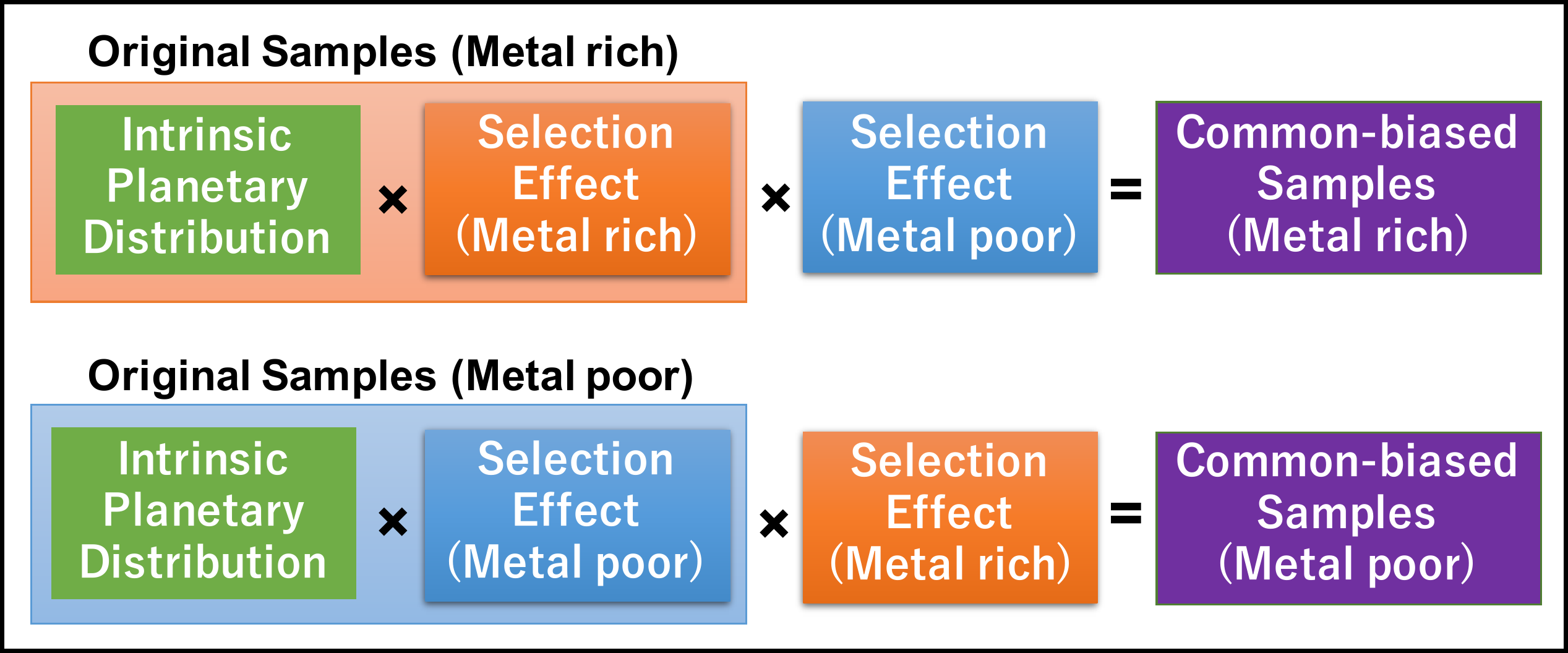}
\end{center}
\end{minipage} \\
\begin{minipage}[t]{1\hsize}
\begin{center}
\includegraphics[width=8.5cm]{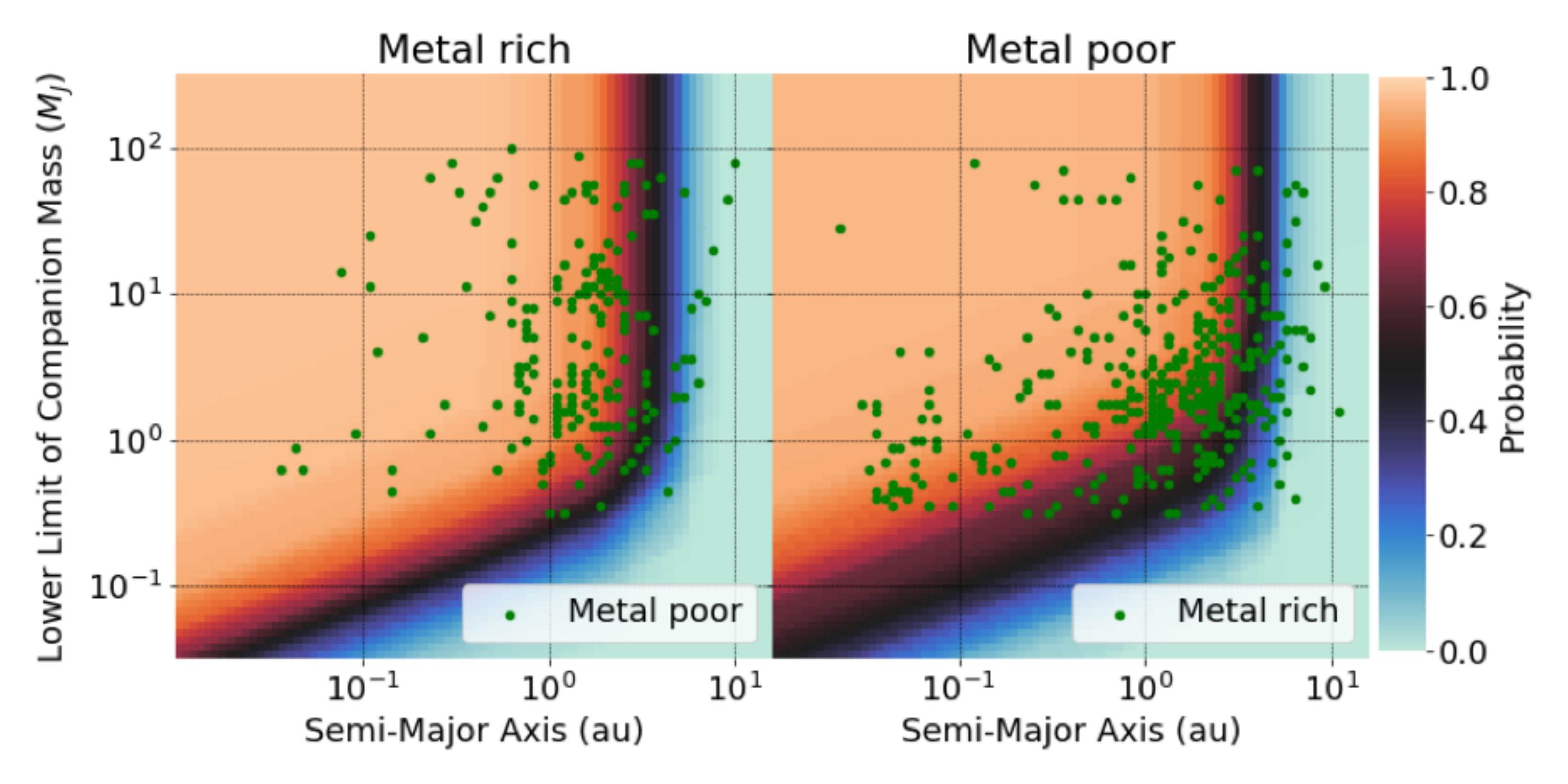}
\caption{$Top:$ Procedure for equalizing the selection biases included in the original samples in the two different metallicity regions. We filtered the original metal-rich (-poor) samples further using the criteria for the radial-velocity measurements for the metal-poor (-rich) original samples. We define the filtered samples so-filtered ``common-biased samples;" i.e., they are biased by common selection effects. This filtering procedure determines the existing probability of each original sample orbiting a metal-rich (-poor) star as the detection probability of the radial-velocity measurements in metal-poor (-rich) region. $Bottom:$ Comparison of distributions of the original metal-poor (left) and -rich (right) samples (green dots) with the detection probability for a companion as derived from the radial-velocity measurements for all stars in the metal-rich (left) and -poor (right), plotted in terms of the companion mass vs. the semi-major axis. For this figure, we applied 0 dex as the metallicity boundary. We defined the probability as the fraction of the number of radial-velocity measurements that can detect a companion to the total number of measurements in each metallicity region.}
\label{fig:procedure}
\end{center}
\end{minipage}
\end{figure}

 Because there is a relation between the planetary-formation processes and the host-star metallicity, as discussed in Section \ref{sec:introduction}, it is preferable that the accuracies and durations of the radial-velocity measurements, used to detect the original samples, be independent of the host-star metallicity. The original samples detected via radial-velocity measurements are influenced by two selection effects: (i) limited sensitivity to long-period planets, owing to the relatively short duration of observations and (ii) limited sensitivity to low-mass planets, owing to insufficient precision in the radial velocity measurements. The maximum semi-major axis $a|_{max}$ and the lower mass limit $M_{p}\sin{i}|_{min}$ of a detectable companion can be determined by the accuracy $\sigma$ in $\rm ms^{-1}$ and the duration $\tau$ in years of the radial-velocity measurements as below \citep{2008ApJ...677.1324T}:
\begin{eqnarray}
\label{equ:a_max}
a|_{max} &\approx& 9.996\times10^{-1}M_{*}^{\frac{1}{3}}\tau^{\frac{2}{3}}\ \ [{\rm au}] , \\
\label{equ:Mp_min}
M_p\sin i|_{min} &\approx& 4.919\times10^{-3}P^{\frac{1}{3}}(1-e^2)^{\frac{1}{2}}M_{*}^{\frac{2}{3}}\sigma\ \ [{\rm M_J}] ,\ \ \ 
\end{eqnarray}
where, $M_*$, $P$, $e$, and $i$ are, respectively, the host-star mass in $\rm M_{\odot}$, and the orbital period in days, eccentricity and orbital inclination of the companion. We have derived the region in which a companion can be detected for each radial-velocity measurement based on Equations (\ref{equ:a_max}) and (\ref{equ:Mp_min}). Focusing on the fact that the distributions of masses and semi-major axes for the original samples discovered in the metal-rich (-poor) region are biased due to the radial-velocity selection criteria for the metal-rich (-poor) stars, we can minimize the impact on the original samples of the difference between the selection effects for the metal-rich and -poor cases by filtering the metal-rich (-poor) original samples with the selection criteria for the metal-poor (-rich) samples. In this procedure, the existing probability of each original sample orbiting metal-rich (-poor) stars is determined by the detection probability of radial-velocity measurements for the metal-poor (-rich) cases shown in Figure \ref{fig:procedure}. This equalizes the selection biases for the samples orbiting the metal-rich and -poor stars.

\begin{figure}[t]
\centering
\includegraphics[width=8.5cm]{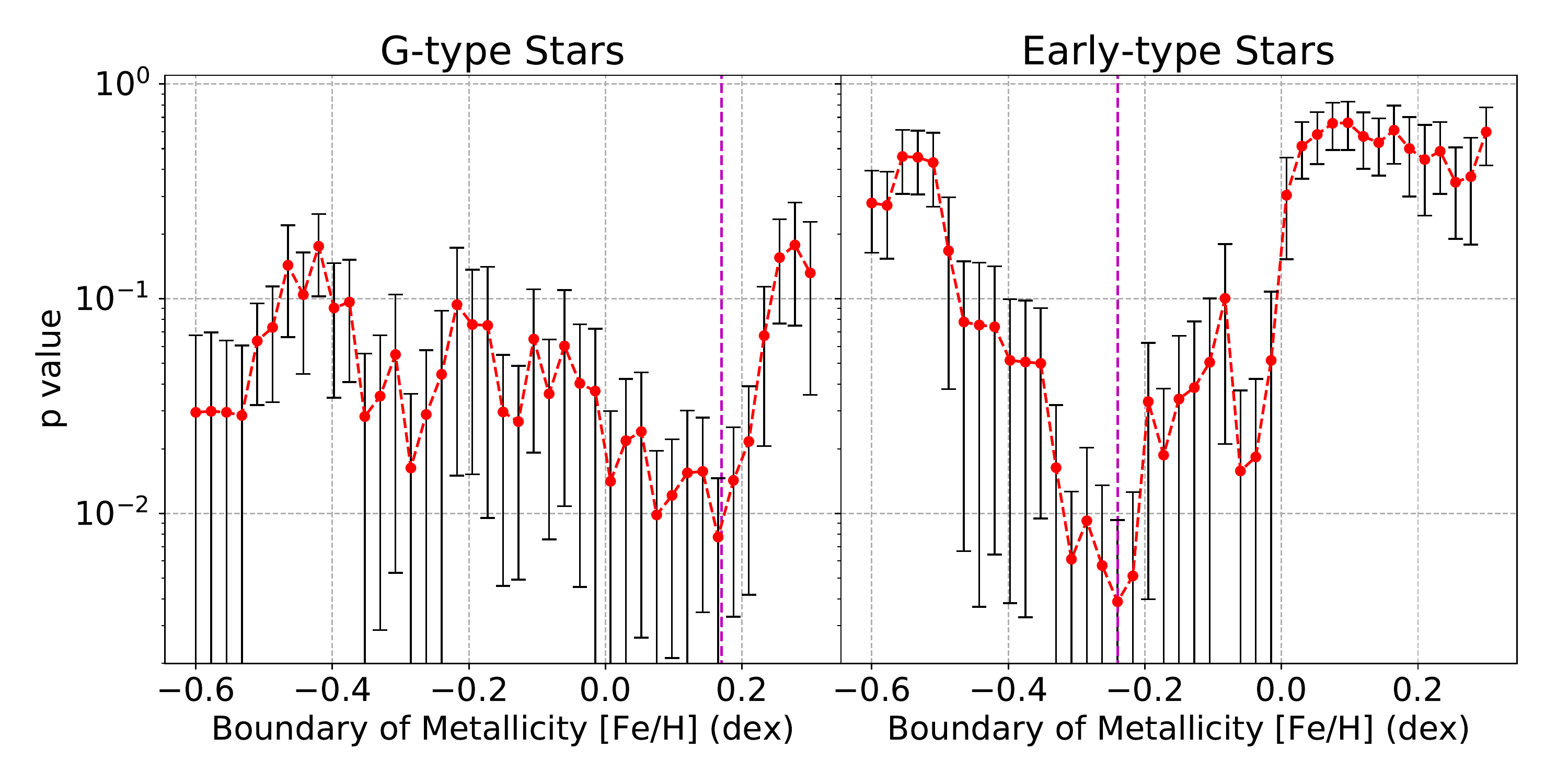}
\caption{The p-values calculated via two-sample Anderson-Darling tests for the lower limit of companion mass of the original samples for G-type stars (left) and for early-type stars (right), as functions of the assumed metallicity boundary. The red points and black vertical bars represent the mean p-values and their standard deviations, respectively. The magenta vertical dashed line represents the host-star metallicity at the lowest p-value for each type star. The number of calculations for each metallicity boundary is 100.}
\label{fig:pvalue}
\end{figure}

Next, we determined the metallicity boundaries that divide the original samples for G-type star and early-type stars into two groups, respectively, such that the distributions of planetary mass in the metal-rich and -poor regions differ the most after taking account of the radial-velocity selection effect. Figure \ref{fig:pvalue} shows the p-values derived by the two-sample Anderson-Darling test for the distributions of the lower mass limits of the common-biased samples for G-type stars and early-type stars, changing the metallicity boundary from -0.6 to 0.3 dex. We iterated the calculation 100 times and averaged the calculated p-values at each divided point to derive the means and standard deviations of the p-values. The minimum p-values from the two-sample Anderson-Darling tests for the distributions of the planetary masses are $7.8\times10^{-3}$ and $3.9\times10^{-3}$ at the metallicities of 0.17 and -0.24 dex, respectively. Therefore, in this study, we used 0.17 and -0.24 dex as the metallicity boundaries for constructing the common-biased samples orbiting G-type stars and early-type stars, respectively.

\begin{figure}
\centering
\includegraphics[width=8.5cm]{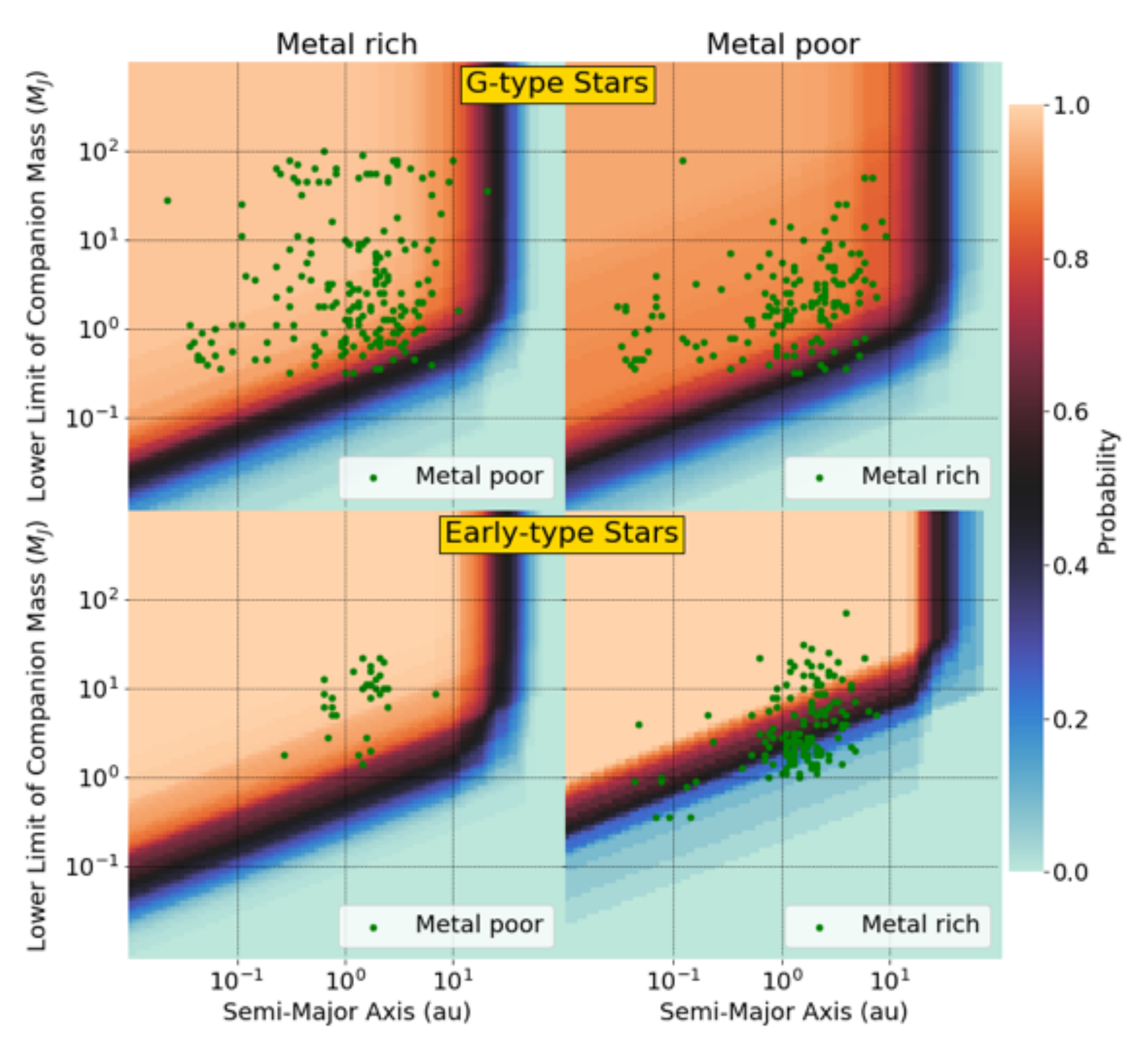}
\caption{Comparison of distributions of companion masses as functions of the semi-major axes for the original metal-poor (left) and -rich (right) samples (green dots) with detection probabilities for a companion as derived from the radial-velocity measurements for G-type stars (upper panel) and for early-type stars (lower panel) in the metal-rich (left column) and -poor regions (right column), plotted in terms of the companion mass vs. the semi-major axis. For this figure, we applied 0.17 and -0.24 dex as the metallicity boundaries for the original samples orbiting G-type and early-type stars.}
\label{fig:bias}
\end{figure}

Figure \ref{fig:bias} compares the detection probabilities for a companion against the radial-velocity measurements for the early-type and G-type stars in both metal-rich and -poor cases, plotted in terms of the semi-major axis and the lower mass limit. As shown in Figure \ref{fig:bias}, while the accuracies of the radial-velocity measurements for the G-type stars do not depend on the host-star metallicities, the accuracies for the metal-poor early-type stars are clearly worse than those for the metal-rich stars. Thus, the radial-velocity selection criteria for the early-type stars depend on the host-star metallicity and affect the distributions of masses and semi-major axes for the two original sub-samples orbiting the metal-rich and -poor early-type stars. 

Finally, according to the determined boundary metallicities for G-type and early-type stars, the detection probabilities for a companion as derived from the radial-velocity measurements for G-type and early-type stars in terms of the companion mass vs. the semi-major axis are constructed (see Figure \ref{fig:bias}). The existence probability of each original sample is determined based on the detection probabilities. In each iteration, each original sample is re-sampled based on the existence probability and the measurement errors of the host-star metallicity, planet mass, semi-major axis, and eccentricity. Note that each measurement error is assumed to follow a normal distribution. We refer to the re-sampled original samples as ``common-biased samples." 

\section{Results} \label{sec:results}

In this section, we explore how many types of extrasolar gaseous objects exist by classifying the common-biased samples with a Gaussian-mixture model. After that, we show the distributions of masses, semi-major axes, and eccentricities for the classified common-biased sub-samples orbiting G-type and early-type stars.

\subsection{Three-Mass Regimes of Gaseous Objects} \label{subsec:mass}

\begin{figure}[t]
\begin{center}
\includegraphics[width=7cm]{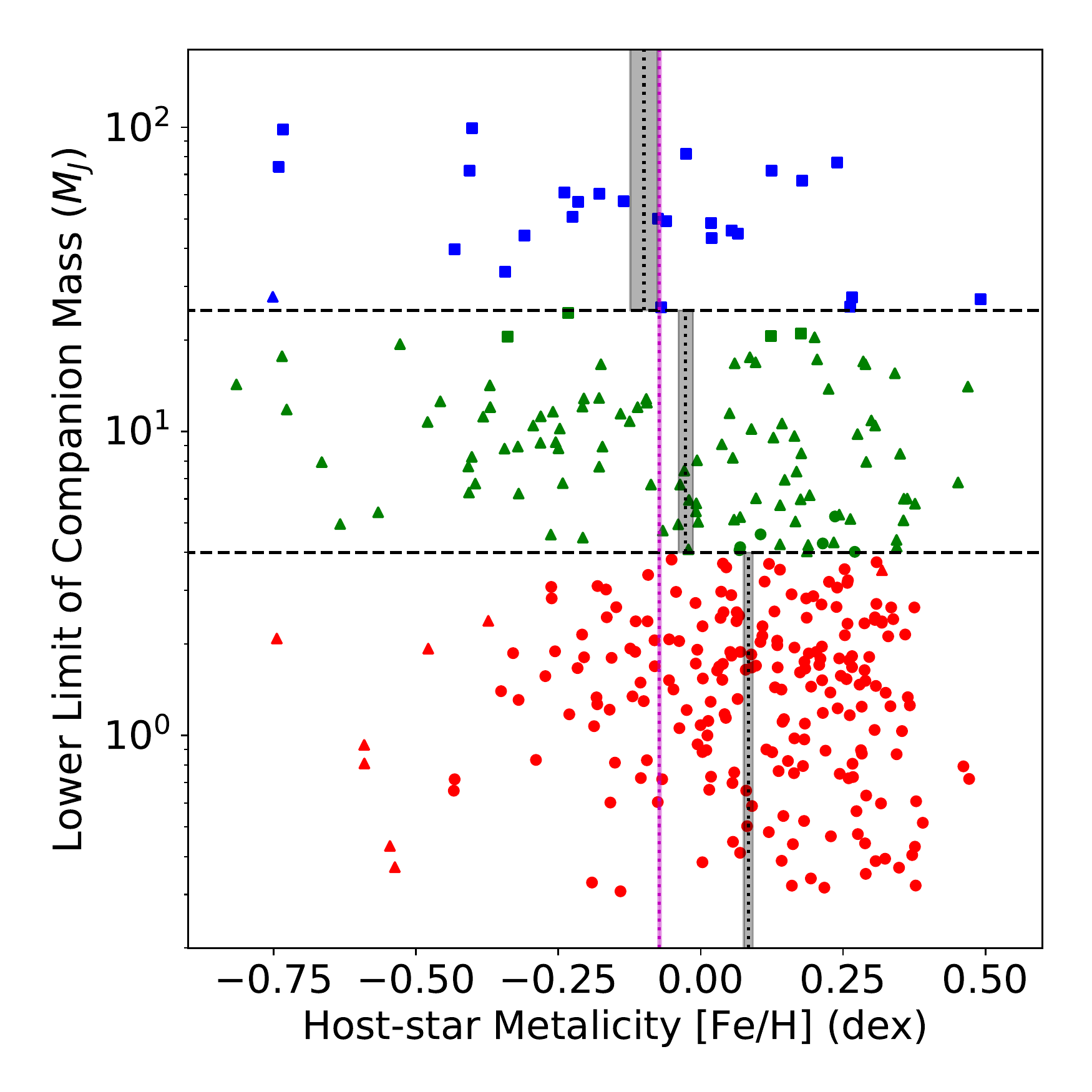}
\caption{Distribution of the lower limits of companion mass as a function of host-star metallicity for the three common-biased sub-samples classified by the best Gaussian mixture model. The different symbols (square, triangle, and circle) represent the classified sub-samples. The three colors of the samples correspond to three mass regimes separated by the two boundary masses of 4 and 25 $\rm M_J$, which are shown by the horizontal long-dashed lines. The vertical short dashed line and gray region in each mass regime represent the mean metallicity and its standard deviation over 1000 iterations, respectively. This distribution of the lower limit of companion mass of the samples $vs.$ host-star metallicity shows one example from among the 1000 calculations. The magenta dashed line and region shows the mean metallicity for all the samples in the Geneva-Copenhagen catalog, where we converted the mean metallicity using the linear regression between the samples in the SWEET-Cat and that catalog. We constructed the common-biased samples for G-type and early-type stars by compensating for the radial-velocity selection effects for the two spectral types of stars, as shown in Figure \ref{fig:bias}.}
\label{fig:gmm}
\end{center}
\end{figure}

We classified the common-biased samples selected from the 569 original ones in a diagram showing the companion mass vs. host-star metallicity, using a Gaussian-mixture model to explore how many distinct sub-samples exist among the extrasolar gas giants discovered to data, given that each sub-sample follows a normal distribution. We evaluated each model using the Bayesian Information Criterion while varying the number of the sub-samples. In this way, we found that a three-component model is the best Gaussian-mixture model for the common-biased samples. Figure \ref{fig:gmm} shows this best-suited model. The common-biased samples are divided into three groups by the two boundary masses of 4 and 25 $\rm M_J$. This three-component model results from a relative paucity of common-biased samples in two specific regions in the diagram of companion mass $versus$ host-star metallicity: the mass range from 20 to 30 $\rm M_J$ around both the metal-rich and -poor stars and the mass range from 0.3 to 4 $\rm M_J$ around the metal-poor stars. As a result, the mean metallicity of the stars hosting gaseous objects with masses between 4 and 25 $\rm M_J$ is lower than that of the samples lighter than 4 $\rm M_J$, and the mean metallicity of the samples more massive than 25 $\rm M_J$ is much lower than those of the other two sub-samples. Thus, we have confirmed that the lower boundary mass is consistent with the results obtained in previous studies \citep{2007A&A...464..779R, 2017A&A...603A..30S, 2018ApJ...853...37S}.　

\subsection{Planetary Distributions around G- and Early-type Stars} \label{subsec:separate}

\begin{figure}[t]
\begin{center}
\includegraphics[width=8.5cm]{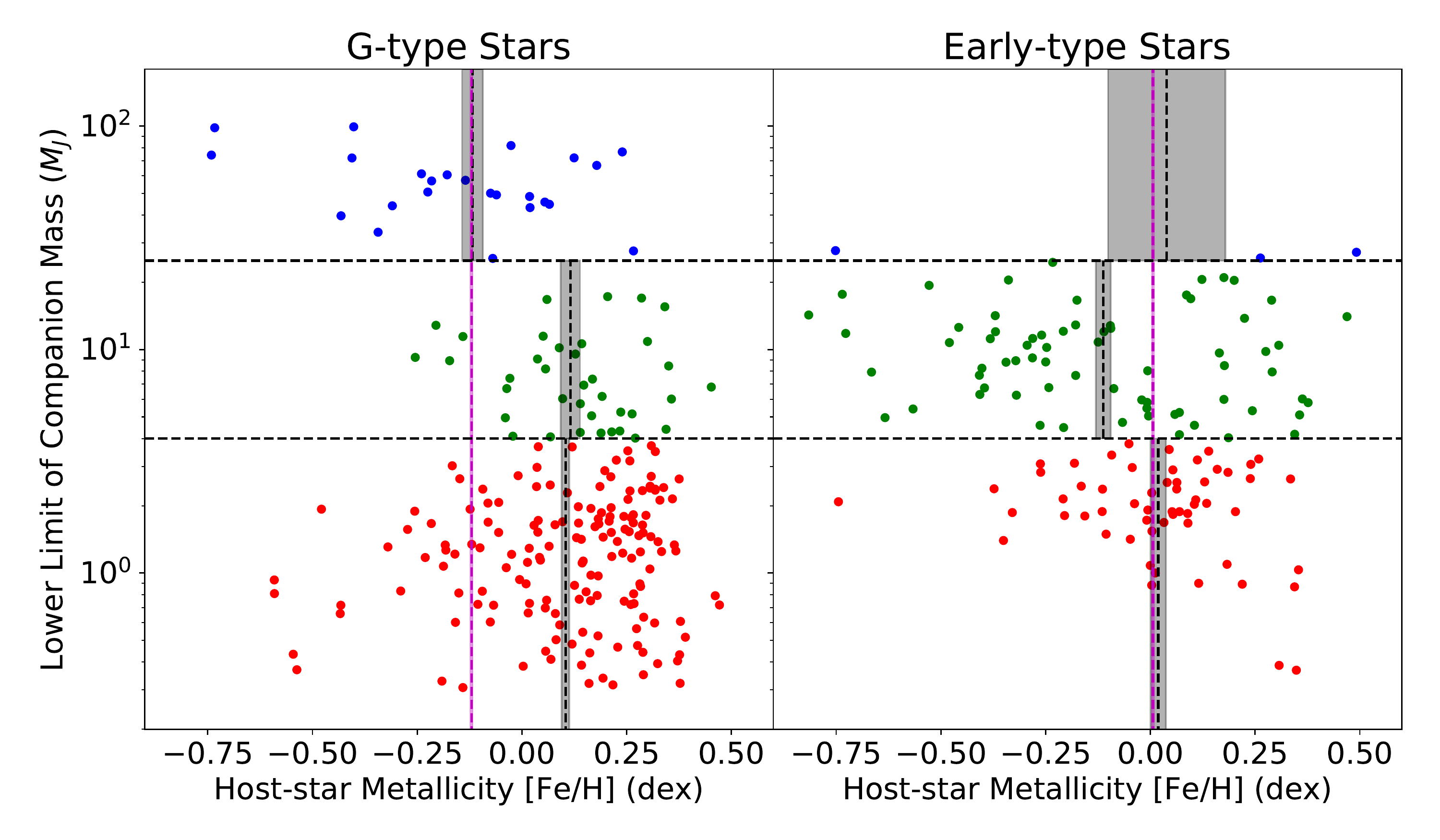}
\caption{Distributions of companion masses as functions of the host-star metallicities for the common-biased samples orbiting G-type stars with masses ranging from 0.8 to 1.3 $\rm M_{\odot}$ (left) and orbiting early-type stars with masses greater than 1.3 $\rm M_{\odot}$ (right). The symbols are same as those in Figure \ref{fig:gmm}.}
\label{fig:metal_Mp_sep}
\end{center}
\end{figure}

Next, we extracted sub-samples from among the common-biased samples orbiting G-type stars with masses ranging from 0.8 to 1.3 $\rm M_{\odot}$ and early-type stars more massive than 1.3 $\rm M_{\odot}$, and we investigated the distributions of host-star metallicities and companion masses around the two types of host-stars. Figure \ref{fig:metal_Mp_sep} shows the distributions of host-star metallicities for the three substellar-mass regimes for G-type and early-type stars. Note that we have constructed the common-biased sub-samples for G-type and early-type stars by accounting for the selection biases for the two different spectral types of stars, as shown in Figure \ref{fig:bias}. 

The mean metallicity of the G-type host-stars with gaseous samples lighter than 25 $\rm M_J$ are much higher than those with samples more massive than 25 $\rm M_J$. This corresponds closely to the metallicity of the nearby G-type stars selected from the Geneva-Copenhagen catalog \citep{2011A&A...530A.138C}. There is also no boundary at 4 $\rm M_J$ in terms of the distribution of the host-star metallicities. The gaseous objects with masses ranging from 4 to 25 $\rm M_J$ -- as well as from 0.3 to 4 $\rm M_J$ -- are thought to be formed via core-accretion. Thus, the upper boundary around 25 $\rm M_J$ apparently reflects the upper mass limit for the core-accreted planets, which is approximately consistent with those found in previous theoretical studies \citep[e.g.,][]{2007ApJ...667..557T, 2009A&A...501.1161M, 2016ApJ...823...48T}. Note that the paucity of samples more massive than 25 $\rm M_J$ around early-type stars also seems to support this upper boundary corresponds as the maximum mass for a planetary object. These results are not in agreement with the conclusion reached by \cite{2018ApJ...853...37S}; there is a transition of the host-star metallicity in the mass range between 4 and 10 $\rm M_J$. However, in the study by \cite{2018ApJ...853...37S}, only seven samples had masses ranging from 4 to 25 $\rm M_J$. In addition, we found that the mean metallicity of the seven samples is 0.12 dex, which is almost equal to that obtained from our analysis. Thus, the mean host-star metallicity for the samples with masses ranging from 4 to 25 $\rm M_J$ around G-type stars is as high as the metallicity for the samples lighter than 4 $\rm M_J$. On the other hand, a star-formation process, such as a gravitational core collapse and the fragmentation of a molecular cloud \citep{2004ApJ...617..559P, 2008ApJ...684..395H}, could possibly form gaseous objects with masses ranging from 5 to 15 $\rm M_J$ \citep{2001ApJ...551L.167B}; an obvious mass boundary may not exist between the planetary-formation and star-formation processes. However, the mean metallicity for the samples much larger than 25 $\rm M_J$ is significantly lower than that for the samples lighter than 4 $\rm M_J$. The star-formation process does not depend on the metallicity of a molecular cloud; therefore, the mass boundary between the star- and planetary-formation mechanisms is thought to be approximately 25 $\rm M_J$.

\begin{figure}[t]
\begin{center}
\includegraphics[width=8.5cm]{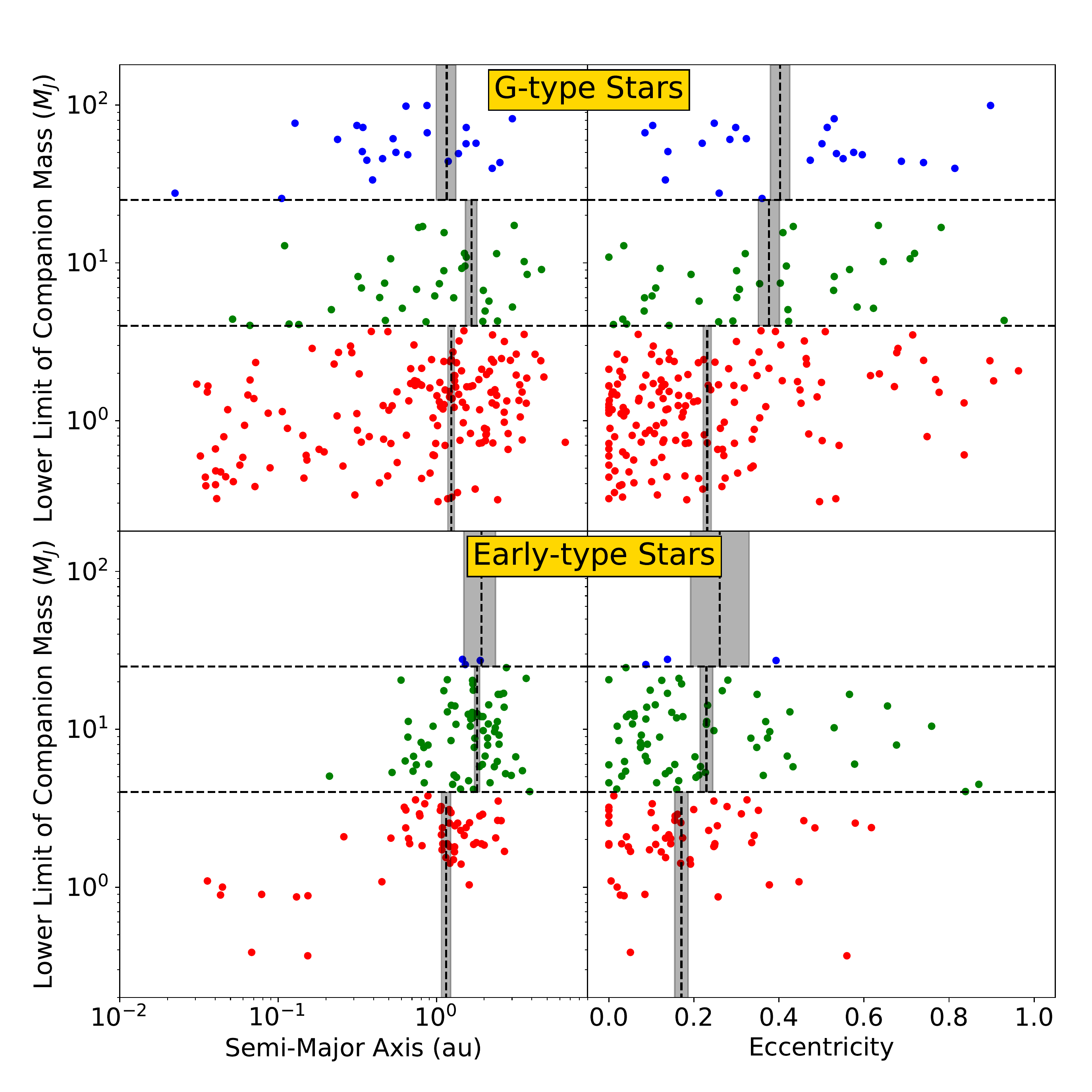}
\caption{Distributions of companion masses as functions of the semi-major axes (left column) and the eccentricities (right column) for the common-biased samples orbiting G-type (upper panel) and early-type stars (lower panel). The symbols and lines are same as those in Figure \ref{fig:gmm}.}
\label{fig:orbit_Mp_sep}
\end{center}
\end{figure}

In contrast, for early-type stars, the mean metallicity for gaseous objects with masses ranging from 4 to 25 $\rm M_J$ is much lower than for the samples lighter than 4 $\rm M_J$. Therefore, the lower boundary mass of 4 $\rm M_J$ seems to indicate that the distribution of host-star metallicities changes significantly at the 4 $\rm M_J$ boundary around early-type stars. This lower boundary also reflects a difference between the planetary formation processes around G-type and early-type stars, since there is no boundary at 4 $\rm M_J$ around the G-type stars. In fact, although the common-biased sub-samples with masses ranging from 4 to 25 $\rm M_J$ orbiting the G-type and early-type stars distribute in the outer region than 0.3 au (Figure \ref{fig:orbit_Mp_sep}), their eccentricity distributions are quite different, as shown in Figure \ref{fig:orbit_Mp_sep}. Note that, because we calibrated 34 and three of the samples, respectively, among the 321 G-type and 164 early-type stars were calibrated in terms of the host-star metallicity, the impact on the distribution of host-star metallicities from non-uniformities in the samples is small. We also emphasize that the number of samples more massive than 25 $\rm M_J$ around early-type stars is much smaller than those around G-type stars. The lack of any samples around early-type stars probably reflects the upper mass limit for objects that evolve from the planetary formation mechanisms. The lack may also show the different formations for sub-stellar components around G-type and early-type stars.

Based on the considerations above, we redefined the samples lighter than 25 $\rm M_J$ as planetary-mass objects and labeled the two sub-samples with masses from 0.3 to 4 $\rm M_J$ and from 4 to 25 $\rm M_J$ as ``intermediate-mass planets" and ``massive planets," respectively. In addition, we re-labeled the samples more massive than 25 $\rm M_J$ as ``brown-dwarfs." Note that the boundary between planetary-mass and brown-dwarf objects established by the deuterium-burning minimum mass of around 10 $\rm M_J$ is semantic \citep{2014prpl.conf..619C}; this boundary has no physical meaning from the evolutionary perspective.

\section{Discussion} \label{sec:discussion}

In this section, first, we discuss the formation process of the intermediate-mass and massive planets orbiting G-type stars, comparing the mass distribution with that predicted by the core accretion model. Then, we focus on the distributions of masses and eccentricities for the intermediate-mass and massive planets orbiting early-type stars. We finally provide an overview of the extrasolar gaseous objects discovered in the context of the two planet-formation scenarios.

\subsection{Planetary Formation Processes around G-Type Stars} \label{subsec:G}

We compared the intermediate-mass and massive planets with simulation data generated by \cite{2012A&A...541A..97M}, who performed population synthesis around a 1 $\rm M_{\odot}$ within the framework of the core accretion model, including planet growth and migration caused by planet-disk interactions. The left panel of Figure \ref{fig:simulation} shows the cumulative mass distributions for the common-biased samples orbiting G-type and early-type stars and the simulation samples. The distribution of the common-biased samples around G-type stars are approximately consistent with that from the simulations. In fact, as shown in the right panel of Figure \ref{fig:simulation}, the mean masses for the observations of planets around G-type stars agree well with the simulation samples over the entire range of metallicities except for the lowest metallicity bin where the population synthesis does not generate any planets. Note that we biased the simulation samples with the selection effects for both the metal-rich and -poor regions, and we restricted the observational samples to the intermediate-mass and massive planets orbiting host stars with masses ranging from 0.8 to 1.3 $\rm M_{\odot}$.

The increases in the eccentricities of the massive planets orbiting G-type stars, shown in Figure \ref{fig:orbit_Mp_sep}, can also be explained by the following two models, which are extensions of the core-accretion model. One extension involves planet-disk interactions at the Lindblad and co-rotation resonances prior to gas dissipation \citep[e.g.,][]{2003ApJ...585.1024G}. According to the numerical simulations performed by \cite{2006A&A...447..369K}, the minimum planet mass necessary to change the disk gas into a high-eccentricity state is 3 $\rm M_J$ for a viscous coefficient of $10^{-5}$. This is approximately consistent with the boundary between the intermediate-mass and massive planets. The second extension involves dynamical instabilities induced by two closely separated gas giants or by three gas giants, so-called "gravitational planet-planet interactions" \citep[e.g.,][]{2013ApJ...775...42I}. The dynamical instability produces a gas giant with an eccentric orbit in the outer region of the planetary system and a circular hot Jupiter in the inner region through tidal circularization \citep[e.g.,][]{1996Sci...274..954R}. In fact, we confirm the paucity of intermediate-mass planets that located within 0.1 AU around metal-poor G-type stars; the gravitational planet-planet interaction is thought to occur only around metal-rich G-type stars. Previous observations have also confirmed that hot Jupiters only orbit metal-rich G-type stars \citep{2013ApJ...767L..24D, 2013A&A...560A..51A}.

The distributions of masses and eccentricities for the intermediate-mass and massive planets orbiting G-type stars are almost explained reasonably well by the core-accretion model. Their distributions support the conclusion that the upper mass limit of the core-accreted planets is around 25 $\rm M_J$. However, a few planets exist around very metal-poor stars located beyond the core-accretion model \citep{2012A&A...541A..97M}. As discussed in Section \ref{sec:introduction}, several discrepancies exist between the planetary distributions revealed by various observations and those predicted by the core-accretion models. Considering that disk instability has an important role in the formation of gas giants around early-type stars (see Section \ref{subsec:early}), some of the massive planets are the disk-instability-induced-planets.

\subsection{Planetary Formation Processes around Early-Type Stars} \label{subsec:early}

\begin{figure}[t]
\begin{center}
\includegraphics[width=8.5cm]{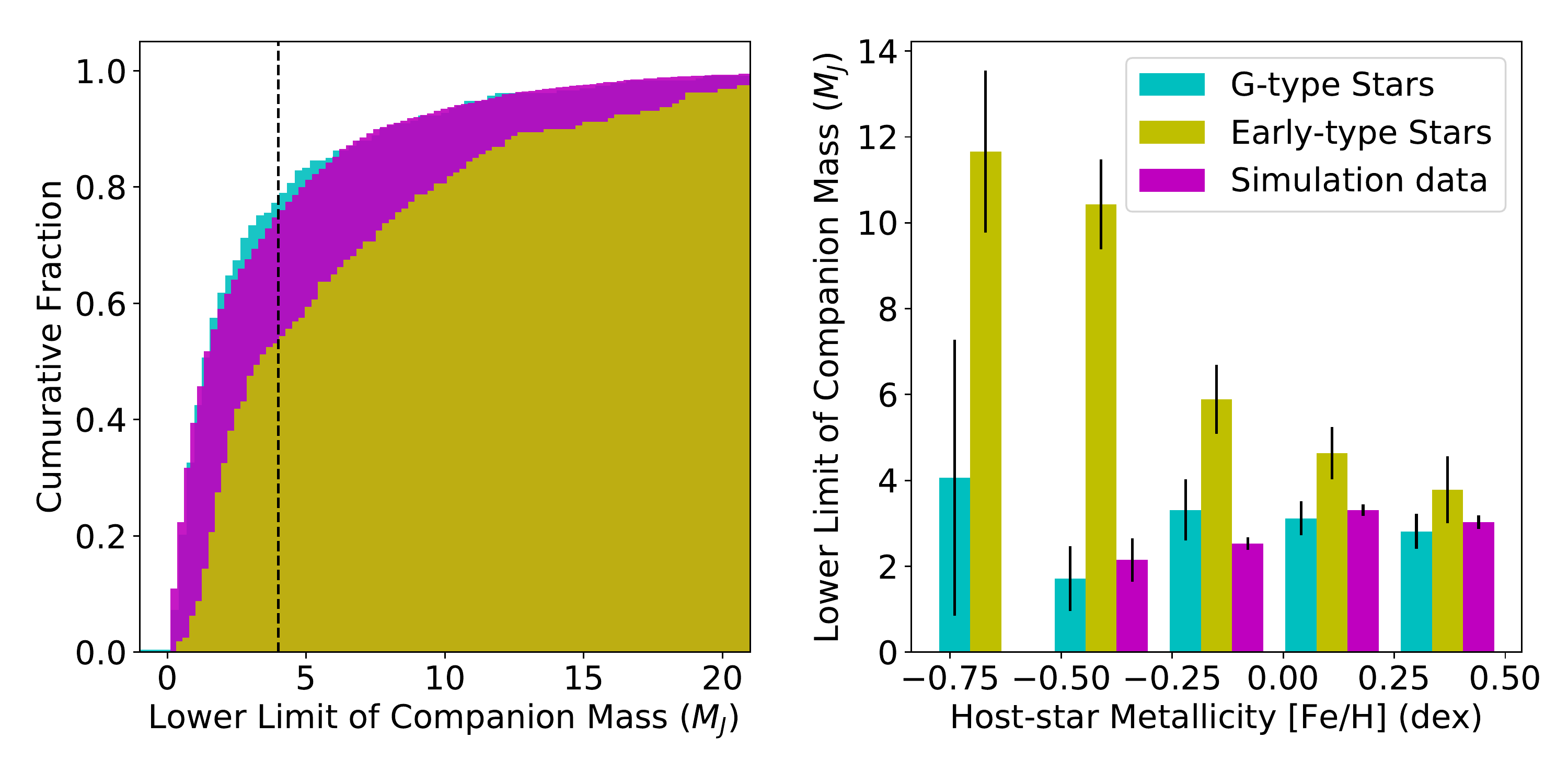}
\caption{$Left:$ Cumulative mass distributions for the common-biased samples with masses less than 25 $\rm M_J$ orbiting G-type stars (cyan bins) and early-type stars (yellow bins). The simulation samples generated by \cite{2012A&A...541A..97M} have companion masses ranging from 0.3 to 25 $\rm M_J$ (purple bins). $Right:$ Histograms of mean masses from the simulation samples (purple bars) and from the common-biased samples with masses less than 25 $\rm M_J$ orbiting G-type stars (cyan bars) and early-type stars (yellow bars).}
\label{fig:simulation}
\end{center}
\end{figure}

The intermediate-mass planets around early-type stars preferentially orbit metal-rich stars, as shown in Figure \ref{fig:metal_Mp_sep}. The intermediate-mass planets orbiting early-type stars as well as the planets around G-type stars are thought to be formed by core accretion. In contrast, the massive planets around early-type stars seem to orbit metal-poor stars preferentially. Note that the mean value of the nearby early-type stars extracted from the Geneva-Copenhagen Catalog may be higher than the true value because of a systematic offset between the Geneva-Copenhagen and SWEET-Cat catalogs. The excess of massive planets orbiting metal-poor early-type stars differs from that expected from the core-accretion theory regarding the following two points: (1) While more-massive planets are expected to be formed around more metal-rich stars \citep{2012A&A...541A..97M}, the mean masses for the intermediate-mass and massive planets orbiting early-type stars clearly increase as the metallicity decreases (Figure \ref{fig:simulation}). (2) In addition, although the mass function for massive planets is theoretically predicted to exhibit a continuous decrease \citep{2009A&A...501.1161M}, the observational samples orbiting metal-poor early-type stars cluster around 4 and 10 $\rm M_J$ (Figure \ref{fig:simulation}). The eccentricities of the massive planets that orbit early-type stars also differ from those around the G-type stars (Figure \ref{fig:orbit_Mp_sep}); the eccentricities of the massive planets around the early-type stars do not seem to be enhanced by planet-disk interactions prior to gas dissipation or by gravitational planet-planet interactions. Thus, the distributions of masses and eccentricities for the massive planets orbiting early-type stars are unlikely to be explained by the bottom-up models.

An explanation for the excess of massive planets orbiting metal-poor stars is that the disk instability acts in the vicinity of metal-poor stars, because a lower mass limit applies to planets formed via the disk-instability mechanism [i.e., corresponding roughly to an order of the Jeans mass \citep{2007ApJ...662.1282M, 2010Wiley}]. Consequently, a sharp increase appears in the planetary mass function around 4 $\rm M_J$. It is also generally accepted that planet formation due to the disk instability tends to occur in the vicinity of metal-poor stars, because the cooling timescale in the disk mid-plane is reduced owing to the low disk opacity \citep{2006ApJ...636L.149C, 2007Arizona}. The low eccentricities of the massive planets orbiting early-type stars are also consistent with numerical simulations \citep{2004ApJ...609.1045M, 2010Wiley, 2011ApJ...731...74B} and with the eccentricities of the four gas giants orbiting HR8799, an A-type star, \citep{2017A&A...598A..83W, 2018AJ....156..192W}. Note that the four gas giants are located in the region beyond the core-accretion model (see Figure \ref{fig:metal_Mp}).

Therefore, the intermediate-mass planets around early-type stars are mainly formed by the core accretion, whereas disk instability plays an important role in the massive-planet formation. However, the upper mass limit for the planetary objects orbiting G-type stars is approximately 25 $\rm M_J$; therefore, some of the massive planets would naturally be core-accreted planets. Both the bottom--up and top--down formation mechanisms occur around early-type stars.

\subsection{Planetary Formation Scenarios} \label{subsec:scenarios}

\begin{figure*}[t]
\begin{center}
\includegraphics[width=17cm]{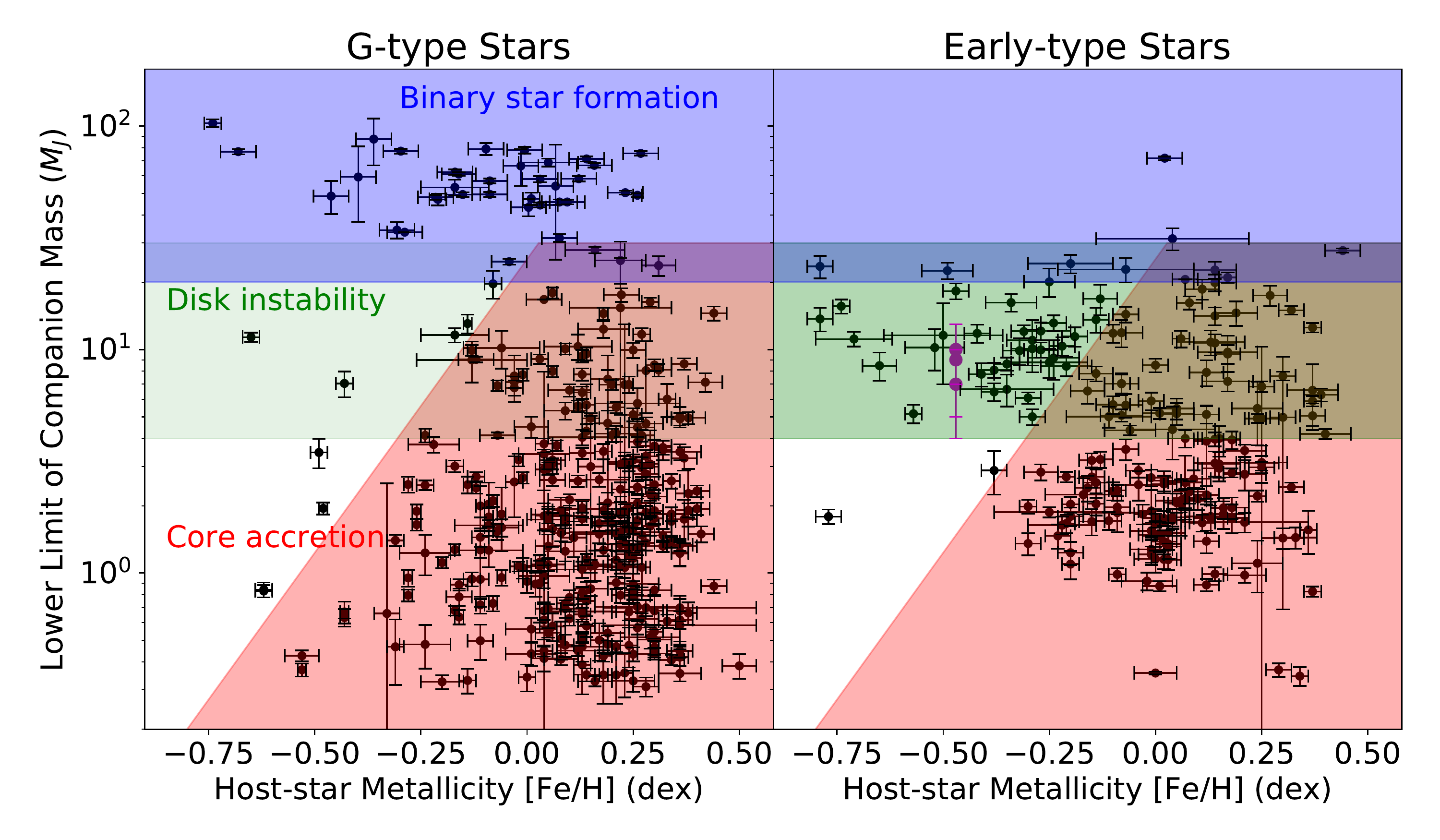}
\caption{Distributions of companion masses as functions of the host-star metallicities for the original samples orbiting G-type stars (left) and early-type stars (right). The original samples (black dots) and the four planets orbiting HR 8799 (purple dots) were compared with expectations from the core-accretion and disk-instability theories. The red, green, and blue regions, respectively, indicate where objects can be formed by core accretion or by disk instability and where binary star formation occurs. The black error bars represent 1 $\sigma$ measurement errors. We have assumed the dependence of the maximum mass on the disk metallicity for core-accreted planets to be same around early-type stars as around G-type stars, which dependence we obtained from the population synthesis performed by \cite{2012A&A...541A..97M}. The masses of the four planets orbiting HR 8799 and the host-star metallicity were obtained from the Extrasolar Planet Encyclopedia catalog \citep{2011A&A...532A..79S}.}
\label{fig:metal_Mp}
\end{center}
\end{figure*}

Based on the considerations above, we have compared the distribution of companion mass as a function of the host-star metallicities for the common-biased samples selected from the 569 original ones with the planet-formation regions expected from the core-accretion and disk-instability models for G-type and early-type stars (Figure \ref{fig:metal_Mp}). Sparsely populated regions of companion mass occur between 20 and 30 $\rm M_J$ around G-type stars and in larger than 25 $\rm M_J$ around early-type stars. The gap in the distribution of massive companions around G-type stars corresponds to the gap between binary-star and planet formation. In addition, the paucity of brown dwarfs around early-type stars may support the conclusion that the boundary mass of about 25 $\rm M_J$ corresponds to the maximum mass of a planetary object. In fact, the upper mass limit is almost consistent with the theoretical expectations \citep{2007ApJ...667..557T, 2009A&A...501.1161M, 2016ApJ...823...48T}. 

\begin{figure*}[t]
\begin{center}
\includegraphics[width=17cm]{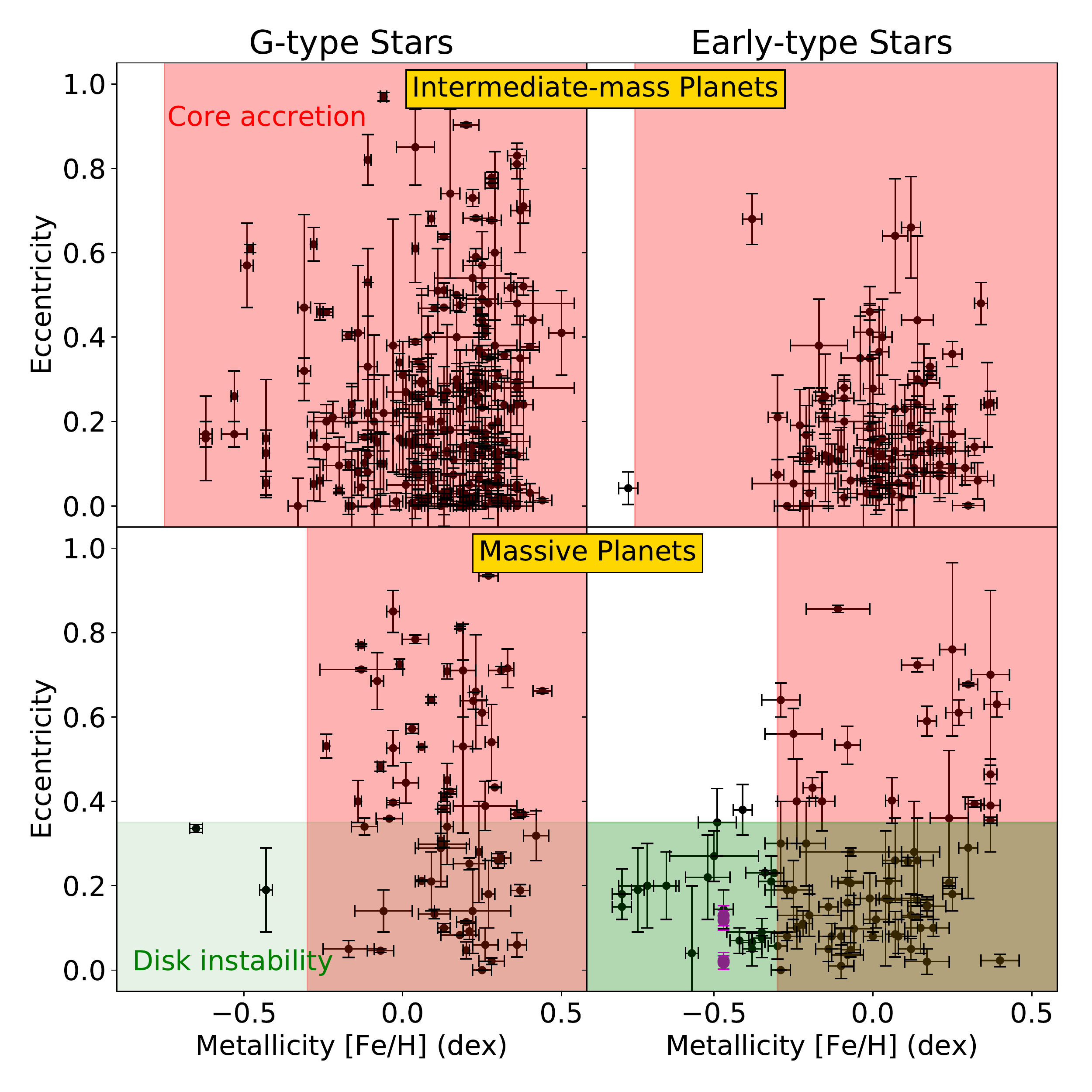}
\caption{Distributions of eccentricities as functions of the host-star metallicities for the intermediate-mass (upper panel) and massive planets (lower panel) orbiting G-type stars (left) and orbiting early-type stars (right). The dots are same as those in Figure \ref{fig:metal_Mp}. The metallicity boundaries of the core-accretion region for the intermediate-mass and massive planets are approximately -0.7 and -0.3, respectively, corresponding to the critical metallicities at 0.3 and 4 $\rm M_J$ for the core-accretion model shown in Figure \ref{fig:metal_Mp}. The upper eccentricity limit for the disk instability was referred to the simulation results \citep{2011ApJ...731...74B}. Note that the upper eccentricity limit is consistent with the other numerical simulations \citep[e.g.,][]{2004ApJ...609.1045M}. The eccentricities of the four planets orbiting HR 8799 were referred to those acquired by the precise astrometric measurements \citep{2018AJ....156..192W}.}
\label{fig:metal_e_sep}
\end{center}
\end{figure*}

The excess massive planets orbiting the metal-poor early-type stars can be explained by the disk instability model, as shown in Figure \ref{fig:metal_Mp}. Comparing the planetary distributions around G-type and early-type stars, the disk-instability-induced planetary formation tends to act more around early-type stars. In addition, when we focused on the distributions of the eccentricities and the host-star metallicities for the samples orbiting early-type stars, we found that the eccentricities of the massive-mass planets orbiting the metal-poor stars are low, whereas the eccentricities of the samples around the metal-rich stars ranged widely from 0 to 0.9; the eccentricity distribution for the massive planets varied largely with the metallicity (see Figure \ref{fig:metal_e_sep}). When the eccentricity of the disk-instability-induced planet is not enhanced \citep{2004ApJ...609.1045M, 2010Wiley, 2011ApJ...731...74B}, the disk instability is thought to mainly occur around the metal-poor stars. In contrast, both the core-accretion and disk-instability mechanisms formulate the wide eccentricity distribution for the samples that orbit the metal-rich stars. The theoretical expectations shown in Figure \ref{fig:metal_Mp} are supported by plotting the eccentricities of the companion masses against the host-star metallicities. Thus, we accepted a hybrid scenario for the planetary formation around G-type and early-type stars.

We found that the variations in the mean host-star metallicities around G-type and early-type stars are approximately 0.25 and 0.13, respectively, corresponding to the factors of 1.8 and 1.3 (see Figure \ref{fig:metal_Mp_sep}). The findings led to the conclusion that the hybrid planetary formation around G-type and early-type stars occurs. On the other hand, the core-accretion population synthesis \citep[e.g.,][]{2004ApJ...604..388I, 2009A&A...501.1139M} strongly depends on the initial parameters, such as the initial disk gas masses and dust--gas ratios of a disk. The disk masses and dust--gas ratios of the core-accretion population synthesis that we adopted ranges from 0.009 to 0.09 $\rm M_{\odot}$ and from 0.04 to 2 times the solar metallicity \citep{2009A&A...501.1139M}, respectively; the two parameters cancel each other in the reproduction of planetary distribution \citep{2012A&A...541A..97M}. It is still not possible to make a uniform prediction of the planetary distribution. The current population synthesis models have difficulty finding slight host-star metallicity variations that could lead to the hybrid planetary formation.

Previous observational studies on dual planetary formation scenarios \citep{2007A&A...464..779R, 2017A&A...603A..30S, 2018ApJ...853...37S} have shown that a boundary mass of 4 to 10 $\rm M_J$ exists in the diagram of host-star metallicities and masses for gaseous objects and have pointed out that this boundary reflects the transition between the two planetary formations; $i.e.$, the upper limit for the core-accreted planets is around 4 $\rm M_J$. However, we found that the boundary at 4 $\rm M_J$ reflects the difference between planetary formations around G-type and early-type stars; the disk instability has a more important role in formation of massive planets around early-type stars. These results have been obtained by a statistical analysis of large-scale samples comprising planets and brown dwarfs that orbit host stars having masses more massive than 0.8 $\rm M_{\odot}$ for each spectral type of stars. The bottom--up and top--down formation mechanisms act in the same planetary mass regimes without these formation mechanisms being divided by a mass boundary.

\acknowledgments
We are sincerely grateful to Dr. Shigeru Ida, Dr. Masahiro Ikoma, and Dr. Kengo Tomida for useful discussion on planetary formation theories. We also thank Dr. Hiroshi Shibai and Dr. Takahiro Sumi for kindly advising us on the method developed for this study. Finally, we express our gratitude to the anonymous referee for significantly enhancing the scientific value of this study.

\appendix
\section{Parameters of orbital parameters of 569 original samples used in this paper}

\begin{longtable}[c]{|l|ccccc|c|}
\caption{Planet parameters}
\label{tab:planet} \\ \hline
Planet & \tablenotemark{1}$a$ (au) & \tablenotemark{1}$M_p\sin i$ ($\rm M_J$) & $K$ (m/s) & $P$ (days) & $e$ & Reference \\ \hline
\endfirsthead \hline \endfoot \endlastfoot \hline
Planet & \tablenotemark{1}$a$ (au) & \tablenotemark{1}$M_p\sin i$ ($\rm M_J$) & $K$ (m/s) & $P$ (days) & $e$ & Reference \\ \hline
\endhead
11 Com b  & 1.2$\pm$0.05 & 16.23$\pm$1.45 & 296.7$\pm$5.6 & 326.03$\pm$0.32 & 0.231$\pm$0.005 & {\cite{2008ApJ...672..553L}} \\
11 UMi b  & 1.89$\pm$0.14 & 16.87$\pm$2.59 & 189.7$\pm$7.15 & 516.22$\pm$3.25 & 0.08$\pm$0.03 & {\cite{2009A&A...505.1311D}} \\
14 And b  & 0.85$\pm$0.03 & 5.0$\pm$0.41 & 100$\pm$1.3 & 185.84$\pm$0.23 & 0.0$\pm$0.0 & {\cite{2008PASJ...60.1317S}} \\
14 Her b  & 2.86$\pm$0.09 & 4.95$\pm$0.3 & 90$\pm$0.5 & 1773.4$\pm$2.5 & 0.369$\pm$0.005 & {\cite{2007ApJ...654..625W}} \\
15 Sagittae b  & 0.35$\pm$0.01 & \tablenotemark{2}68.7$\pm$2.75 & - & 73.3$\pm$3.05 & 0.5$\pm$0.01 & {\cite{2012ApJ...751...97C}} \\
16 Cyg B b & 1.69$\pm$0.04 & 1.7$\pm$0.11 & 50.5$\pm$1.6 & 798.5$\pm$1.0 & 0.681$\pm$0.017 & {\cite{2006ApJ...646..505B}} \\
18 Del b  & 2.35$\pm$0.08 & 8.52$\pm$0.59 & 119.4$\pm$1.3 & 993.3$\pm$3.2 & 0.08$\pm$0.01 & {\cite{2008PASJ...60..539S}} \\
24 Boo b  & 0.27$\pm$0.01 & 1.79$\pm$0.13 & 59.9$\pm$3.25 & 30.35$\pm$0.01 & 0.042$\pm$0.038 & {\cite{2018PASJ...70...59T}} \\
\hline
\end{longtable}
\tablenotetext{1}{The semi-major axis and companion mass were calculated based on Equations (\ref{equ:a}) and (\ref{equ:Mp}).}
\tablenotetext{2}{The lower limit of companion mass was applied as that listed in reference.}

\renewcommand{\thetable}{\Alph{table}}
\begin{longtable}[c]{|l|cccc|c|}
\caption{Parameters of stars hosting 569 original samples and parameters of radial-velocity observations used in this paper}
\label{tab:star} \\ \hline
Host-star & [Fe/H] (dex) & $M_{*}$ ($\rm M_{\odot}$) & $\sigma$ (m/s) & $\tau$ (years) & Reference for $\sigma$ and $\tau$ \\ \hline
\endfirsthead \hline \endfoot \endlastfoot \hline
Host-star & [Fe/H] (dex) & $M_{*}$ ($\rm M_{\odot}$) & $\sigma$ (m/s) & $\tau$ (years) & Reference for $\sigma$ and $\tau$ \\ \hline
\endhead
11 Com  & -0.34$\pm$0.06 & 2.14$\pm$0.28 & 25.5 & 3.1 & {\cite{2008ApJ...672..553L}} \\
11 UMi  & -0.13$\pm$0.04 & 3.4$\pm$0.76 & 27.5 & 3.5 & {\cite{2009A&A...505.1311D}} \\
14 And  & -0.29$\pm$0.03 & 2.38$\pm$0.29 & 20.3 & 4.1 & {\cite{2008PASJ...60.1317S}} \\
14 Her  & 0.38$\pm$0.04 & 0.99$\pm$0.09 & 13.0 & 7.1 & {\cite{2007ApJ...654..625W}} \\
15 Sagittae  & 0.05$\pm$0.07 & 1.08$\pm$0.04 & \tablenotemark{b}- & \tablenotemark{b}- & {\cite{2012ApJ...751...97C}} \\
16 Cyg B & 0.09$\pm$0.01 & 1.01$\pm$0.08 & 7.3 & 16.0 & {\cite{2006ApJ...646..505B}} \\
18 Del  & 0.0$\pm$0.03 & 1.76$\pm$0.18 & 15.4 & 4.8 & {\cite{2008PASJ...60..539S}} \\
24 Boo  & -0.77$\pm$0.03 & 2.72$\pm$0.21 & 26.51 & 13.2 & {\cite{2018PASJ...70...59T}} \\
\hline
\end{longtable}
\tablenotetext{a}{The host-star metallicity was calibrated based on linear regression derived from the host-star metallicity correlations between the SWEET-Cat and Geneva-Copenhagen catalogs.}
\tablenotetext{b}{The accuracy and duration of the radial velocity measurement for the sample were not used for constructing the detection probabilities for a companion as derived from the radial-velocity measurements.}

\end{CJK*}
\end{document}